\documentstyle[aps,floats,prl]{revtex}
\begin{document}
\title{A mystery of conformal coupling}

\author {E.~A.~Tagirov,}
\address
{N N Bogoliubov Laboratory of Theoretical Physics, Joint Institute for Nuclear
Research, Dubna, 141980, Russia,\\
    e--mail: tagirov@thsun1.jinr.ru}
\maketitle


\begin{abstract}
An origin and necessity  of so called conformal  (or,
Penrose-Chernikov-Tagirov) coupling of scalar field to metric of
$n$-dimensional Riemannian space-time is discussed in brief. The corresponding
general-relativistic field equation  implies a one-particle (quantum
mechanical) Schr\"{o}dinger Hamiltonian which depends on $n$, contrary to the
Hamiltonian constructed  by quantization of geodesic motion, which is the same
for any value of  $n$. In general, the Hamiltonians can coincide only for $n =
4$, the dimensionality of the ordinarily observed Universe. In view of the
fundamental role of a scalar field in various cosmological models,
 this fact may be of interest for models of  brane worlds where $n > 4 $.\\


\end{abstract}

\medskip

\newcommand{\defst}{\stackrel{\rm def}{=}}
\newcommand{\vp}{\varphi}
\newcommand{\rinf}{V_n}
\newcommand{\rin}{$V_n$}

\bigskip
 In papers \cite {cht,tg1},  it  was found  that, in the
framework of the canonical quantum field theory (QFT),  wave equation
determining quantum motion of a neutral scalar particle in the general
$n$-dimensional Riemannian space-time $V_n$  with the metric form
\begin{equation}
 ds^2 = g_{\alpha\beta} (x) dx^\alpha dx^\beta,
 \quad \alpha, \beta, ... = 0, 1, ..., n-1,
\end{equation}
should {\it necessarily} be of the form
\begin{equation}
 \Box_g\varphi + \xi\, R(x)\, \varphi +
\left(\frac{mc}{\hbar}\right)^2  \varphi =  0. \label{r}
\end{equation}
Here $\varphi \equiv \varphi(x)$ is  the real scalar field, $\Box_g $ is the
ordinary generalization  of the d'Alembert operator to $V_n$, i.e.
$$\Box_g \defst \frac1{\sqrt{-g}} \frac\partial{\partial x^\alpha}
\left(\sqrt{-g} g^{\alpha\beta} \frac\partial{\partial x^\beta}\right),
 \quad  g \defst {\rm det}\|g_{\alpha\beta}\|, $$
 $R(x)$ is the scalar curvature of  $V_n$, and
 \begin{equation}
\xi = \xi_c (n) \defst \frac{n-2}{4(n-1)}\ .
\end{equation}
This equation is  referred as  the equation of with \emph{conformal coupling}
(CC) to external gravitation or  as  the Penrose--Chernikov--Tagirov (PCT)
equation. If  $m = 0$, the equation is \emph{covariant} with respect to the
 \emph{conformal mapping} of  manifold $V_n$ to $V'_n$  simultaneously with
 $\varphi$ :
\begin{equation}
V_n \rightarrow V'_n : g_{\alpha\beta}(x)\ \rightarrow  \ g'_{\alpha\beta}(x)
= \Omega^2 (x) g_{\alpha\beta}(x),\\
\varphi(x)\ \rightarrow  \ \varphi'(x)=
  \Omega^\frac{2-n}{2} \varphi(x) \label{cm}
\end{equation}
and is \emph{invariant} with respect to \emph{conformal transformations of
coordinates} $x^\alpha$ if $V_n$ admits such transformations which is not the
generic case, see  more  details on the conformal symmetries, e.g., in
\cite{tt}. Thus, the CC equation (1) with $\xi = \xi_c $ and $ m = 0$ has the
same conformal symmetry as the wave equations for a photon and neutrino and
the equation of isotropic geodesic lines (the classical equation of motion of
massless particles) have. This symmetry feature of CC equation for $ n =4$,
i.e., for $\xi_c = 1/6 $, had been noted by R. Penrose \cite{pen} but only as
a mathematical possibility, with no discussion and physical consequence.

However,  eq.(1)  violates explicitly the well-known Einstein  principle  of
local equivalence  according to which  it should take  the form of the
ordinary Klein-Gordon equation  at the  origin   of the normal Riemannian
(locally quasi-Lorentzian) space-time coordinates  fixed  at a given point
$\{x^\alpha\}_0\, \in \,V_n$, see below. The latter requirement is satisfied
generally only if $\xi = 0$, that is  for the minimal coupling (MC) of $\phi$
to gravitation. In itself, the conformal symmetry mentioned above was not   a
sufficient argument in 1960's to break, in favor of CC , the fundamental
principle sanctified by the name of the originator of General Relativity. The
main arguments of \cite{cht,tg1} were based not on the conformal symmetry in
the particular case of  $m = 0$; they resulted from the physical idea that
motion of a high-energy quantum particle  is asymptotically  close to  that of
the classical one and, thus, to the geodesic lines  in the case of $V_n$. In
view of the radical change which is caused by these arguments  and their
importance for  the final conclusion of the present paper, let us consider in
brief the logic of which lead us (N. Chernikov  and the present author) to
$\xi = \xi_c$.

 We considered canonical quantization of the field in the
 $n$-dimensional  de Sitter space-time $dS_n$ of  radius $r$ and,
 at first, had naturally started with
 the MC version of  eq.(1), i.e with $\xi = 0$,  in accord with
 the principle of equivalence. We looked for a Fock space the
   cyclic state of which  is invariant with respect to
   the de Sitter  group $SO (1, n)$ (the symmetry of $dS_n$) and
   the subspace  of the \emph{one-quasi-particle} states  of which realizes
   an irreducible
 representation of this group.   As a result, we found  an one-parameter
 family of  such Fock spaces. They  are  unitarily non-equivalent for
 different values  of the parameter. The question was: is there a
 physical reason to   distinguishes  one  of them?
Our reasoning  was that such  representation space   should include states
corresponding to the quasi-classical motion of \emph{a particle}. The phases
of such states should asymptotically satisfy the Hamilton-Jacobi equation for
large eigenvalues of the  Casimir operator of the representation of
 $SO_{1,n}$. However, it had turned out that  no such representation space
 exists. Instead, we observed that such space  would exist and be unique if
 the mass term $(mc/\hbar)^2 \phi $ in  the initial MC version of eq. (1)
 were shifted by $(n(n-2)/(4 r^2)) \phi $.
We had just done it and realized  that this term  is just the value of the
second term of eq.(1) for $dS_n$ if $\xi = \xi_c (n) $.

Further,  it is  easy  to see that  $R(x)$ is a unique scalar of the dimension
${\rm [length]}^{-2}$ constructed of  the metric tensor and its derivatives in
the generic $V_n$, which takes on the value $ n(n-1)/r^2$ in $dS_n$
 \cite{tg1}. It should be noted also that, in the
Friedmann--Robertson--Walker models, the  approach of \cite{cht} singles out a
unique family of unitary equivalent  Fock spaces \cite{bt}.

As soon as eq.(1) is  accepted with $\xi \neq 0$ (this case,  more general
than that  of  $\xi = \xi_c$,  is appropriate to refer as the non-minimal
coupling (NMC) version of eq.(1) ), a very important physical consequence
follows. Namely, variation of the corresponding action integral by
$g^{\alpha\beta}(x) $ gives a new energy-momentum tensor
 $T_{\alpha\beta}(x;\, \xi)$  which differs from the traditional  one
$T_{\alpha\beta}(x; 0)$   for  the minimal coupling:
\begin{equation}
T_{\alpha\beta}(x; \xi)  =  T_{\alpha\beta}(x; 0) \\
-  \xi (R_{\alpha\beta } - \frac12 R g_{\alpha\beta } +  \nabla_\alpha
\partial_\beta - g_{\alpha\beta }\Box_g ) \phi^2,  \label{t}
\end{equation}
where   $\nabla_\alpha$  is the covariant derivative and
   $T_{\alpha\beta}(x;0)$ is the energy-momentum tensor.
 If $\xi = \xi_c(n) $ ,   one has
\begin{equation}
T^\alpha_\alpha (x; \xi_c )  =  \left(\frac{mc\phi}{\hbar}\right)^2
\end{equation}
\emph{ on solutions of}  eq.(1). Thus, the trace of $T_{\alpha\beta}(x; \xi)$
vanishes  for $m = 0$ contrary to the one of $T_{\alpha\beta}(x; 0)$. The
vanishing trace is necessary  for  definition of the conserved quantities of
massless fields, which  correspond to the conformal symmetry in $V_n$'s having
such a symmetry, see details in \cite{cht}. Here, the most remarkable fact is
that the difference between $T_{\alpha\beta}(x; \xi)$  and $T_{\alpha\beta}(x;
0)$ retains even in the flat space-time $E_n$, where $R(x)\equiv 0$ and it is
of no importance which value  the constant $\xi$ has. In particular, this
result eliminated  a certain  perplexity  existed until then and consisted in
that the conformal invariance of the ordinary d'Alembert equation
 $\Box\phi = 0 $ does not lead to the standard  expressions for the corresponding
conserved quantities in terms of the canonical energy-momentum tensor contrary
to the Maxwell, Dirac-Weyl and isotropic geodesic world-lines equations. It
was known (private communication by  V.I. Ogievetsky) that  this paradoxical
situation can be resolved by  some extra terms to the canonical tensor but but
an origin of them was not known.

  Shortly later, it was also shown \cite{ccj}  that "the new improved
  energy-momentum tensor"
 has  finite  matrix elements in  the ordinary \emph{Poincare-invariant}
 quantum theory of  the non-linear scalar field in $E_4$  with
 self-interaction  $ \lambda \phi^4$ (in the Lagrangean).

As concerns the principle of equivalence, a more deep look at eq.(1) taking
into account  its quantum  nature shows its actual accord with the principle
of equivalence formulated in terms of the Feynman propagator \cite{grb}.

 By now,  eq.(1)  came to the ordinary and wide use for the scalar
field in $V_n$. The  most models of cosmic inflation include  a fundamental
inflaton  scalar field and there are serious arguments that it should
necessarily obey the NMC equation, very probably with  the conformal covariant
self-interaction term \cite{far}. Now, even  prospectives of detection of the
value of $\xi$ from astrophysical observational data are under discussion,
see, e.g., \cite{prkp} and references therein.

In view of the importance of the question, it   should be emphasized that $\xi
= \xi_c$ had arose in the process of determination of the one-particle
subspace of the particle-interpretable Fock space, that is in the process of
extraction of  quantum mechanics (QM) of a particle from the QFT in $dS_n$.
 This is a particular case of that which may be called the field--theoretical
 (FT) approach to construction of  QM in $V_n$; development of the approach to
  the general $V_n$ is given in \cite{tg2,tg3}.

  However, there is another
 approach to the same problem: quantization of the
corresponding classical mechanics (CM). For the free scalar  particle, the
latter is  the Hamilton theory of  the geodesic lines in $V_n$. Which value of
$\xi$ does this approach  suggest? There are   different   formalisms  of
quantization  of the finite-dimensional Hamiltonian  dynamics'.  Those of them
which  reproduce, under appropriate conditions, the mathematical structure of
the standard non-relativistic QM require \emph{logically}  $\xi = 1/6$ for
\emph{any} value of dimensionality $n$ of $V_n$ ! That is the values of $\xi $
to which  lead the two approaches coincide only if $n = 4$, the space-time
dimensionality of the Universe which we observe. This conclusion is so
surprising that  some explanations are necessary  on the way which leads to
it.

In fact,   one should  confront the two approaches only on the level of the
standard non-relativistic  quantum mechanics'   which follow from them. To
this end, a particular frame of reference should be introduced, that is a
representation of $V_n$  as a foliation  by space-like hypersurfaces $\Sigma_t
(x) = const $ (spatial sections of $V_n$) enumerated by values of an evolution
parameter $t$, i.e. the time coordinate. Since  a value of the constant $\xi$
only is in interest,  the task can  be simplified essentially by restriction
to the case of the globally static $V_n$ and the normal geodesic frame of
reference in which the metric  has the form
\begin{equation}
 ds^2 = c^2 dt^2  -  \omega_{ij} (u) du^i du^j, \quad
 u\in \Sigma_t,
\end{equation}
where  $i, j, ... = 1,..., n-1$.  It is convenient that  scalar curvatures
$R(u)$ for the metric tensors $g_{\alpha\beta}(u)$ and
 $\omega_{ij}(u) $ coincide in the globally static case.

 If the mentioned assumptions are made, the systems under consideration
(the scalar field and the particle moving along a geodesic line) have
conserved energy. Therefore the vacuum and quantum  particle states can be
determined, which are stable. In the both approaches, the quantum one-particle
state space can be represented as the space  of functions $\psi (t,\, u )$
which are square integrable (as functions of $u$) over $\Sigma_t$ with its
invariant measure and therefore are the probability amplitudes  to detect the
particle at the point $u$ of $\Sigma_t$. They are solutions of the following
Schr\"{o}dinger equation \cite{tg2,tg3}:
\begin{equation}
  i\hbar \frac{\partial}{\partial t} \psi
  =   mc^2 \sqrt{1   + \frac{2\hat H_0}{mc^2}}\   \psi, \label{ham}
\end{equation}
 $\hat H_0$ being the non-relativistic  Hamiltonian
\begin{equation}
\hat H_0 = - \frac{\hbar^2}{2m}(\triangle_\Sigma - V^{(q)}(u)),
\end{equation}
 where $\triangle_\Sigma$ a  is the Laplace--Beltrami operator on $\Sigma_t$
 and  $V^{(q)} (u) $ is the so called  \emph{quantum potential}.

 In  the FT--approach, where eq.(\ref{ham}) arises as a result of restriction
 to the positive energy solutions of
 eq.(1), see \cite{tg2},   one has  $V^{(q)} (u) = - \xi R(u)$  with
 $\xi = \xi_c$ to satisfy  the requirements of \cite{cht,tg1}.

In the CM-approach, the situation is more complicate,  since the function
 $V^{(q)} (\xi)$ depends not only  on a formalism of quantization, but also on
 choice of coordinates $u^i$, that  is  $V^{(q)} (u)$ is not a scalar. %
 This circumstance looks very strange, but in  the CM--approach
 coordinates $u^i$   together with  the canonically conjugate momenta  $p_i$
 are  the basic \emph{observables} contrary the FT-approach where the
 corresponding
 observables are quadratic functionals  of the field \cite{tg2}.
According to  analysis by C. Rovelli \cite{rov},  information on quantum
 system  unavoidably includes some information on the classical devices
  which are  used  to observe the system and one may think that  the
 quantum potential includes information on the system of coordinates used
 \emph{to observe} a position of the particle.
 Therefore,  it is not so suprising  that quantum  dynamics
 depends on which system of  the basic observables $u^i,\ p_i$  is taken to
 describe it.

 Then, to compare  quantum potentials arising in different formalisms of
 quantization in
 the  CM-approach    a concrete system of coordinates $u^i$ should be
 chosen.  Such system is suggested  by  B. DeWitt's
 construction \cite{dw}  of  the WKB-propagator in $V_n$, which is
   equivalent, in  fact, to introduction of  \emph{Riemannian} coordinates
$y^i $ in a  neighborhood of  a point of  observation $y^i_0 = y^i(u_0)$, see
details in \cite{tg3}. They define a position  of the point  $u$ through the
geodesic distance $s(u,\, u_0 )$ and the unit tangent vector $(du^i/ds)_0 $
along the geodesic line at $u_0$ :
\begin{equation}
y^i (u) \defst s(u,\,u_0)\left(\frac{du^i}{ds}\right)_0 \label{y}
\end{equation}
In this notation, the quantum potential $(\hbar^2/2m) V^{(q)} (y)$ in the
Schr\"{o}dinger equation (\ref{ham}) corresponding to DeWitt's propagator has
the form
\begin{equation}
   V^{(q)}(y) = \frac16 R(y) + O(y^i - y_0^i).      \label{vr}
\end{equation}
Thus, in this form, the non-minimal term in the non-relativistic Hamiltonian
appeared as early as 1957. However, DeWitt considered it as an unfavorable
phenomenon and preferred to avoid  it  changing the Lagrangean in the action
integral. This curious story shows once more the  radical nature of transition
to conformal coupling in \cite{cht}.

Further, it was shown in  paper \cite{tg3} that one can  use ambiguities in
the Beresin--Shubin \cite{bsh} and Feynman quantizations  of the geodesic
motion so that the quantum potential will be of the form (\ref{vr}) when  the
Riemannian coordinates are taken as observables of space position. This
condition removes the ambiguities and fixes these formalisms in their
application to the geodesic motion as well as establishes their concordance
with the WKB formalism. However, the potential (\ref{vr}) does not depend on
$n$, the dimensionality of $V_n$. This is consistent
 with the field-theoretically deduced $V^{(q)} (u) = -\xi_c (n) R(u)$ only if
 $n=4 $. This fact is just that was meant above as logical inconsistency
 of the FT- and CM-approaches  to formulation of quantum mechanics in $V_n$.
Attempts to   explain   it by  topological  difference between  $dS_n$ and
 $V_n$  (in the simple versions of
 the CM approach \cite{tg3}, the latter  is supposed to be topologically
 trivial) or by that the two approaches should not be  compatible because
 only one of them is
 correct will meet the question: then, why they are compatible for
 $n =  4$?

A pragmatically  inclined physicist might be satisfied with that the  matters
are OK, at least, in the  four-dimensional Universe observed ordinarily  and
consider the problem with $n \neq 4 $ as having only an academic interest.
However,
 recent time  very interesting    models  of "the world on a brane" are in
 intensive discussion, according to which  the
 ordinary matter  "lives" (evolves in time)  on a three dimensional space
 while  gravitation and, probably, some other exotic fields  act in  an
 embracing  space of  a higher dimensionality, see a  review of these models,
 e.g., in \cite{rub}. Is a scalar field, which plays a fundamental role in
 various  cosmological models, an "ordinary" matter or not, this is a question
 of the concrete model.

\end{document}